# The Qitai Radio Telescope

Na Wang[1,2,3]*, Qian Xu[1,2,3], Jun Ma[1,2,4], Zhiyong Liu[1,2,3], Qi Liu[1,2,4], Hailong Zhang[1,2], Xin Pei[1,2,4], Maozheng Chen[1,2,4], Richard N. Manchester[5], Kejia Lee[6], Xingwu Zheng[7], Hans J. Kärcher[8,10], Wulin Zhao[9], Hongwei Li[9], Dongwei Li[9], Martin Süss[10], Matthias Reichert[10], Zhongyi Zhu[11], Congsi Wang[12], Mingshuai Li[1,2], Rui Li[1,2], Ning Li[1,2], Guljaina Kazezkhan[1,2], Wenming Yan[1,2,3], Gang Wu[1], Lang Cui[1,2,3], Ming Zhang[1,2,3], Haitao Li[1]

[1] *Xinjiang Astronomical Observatory, Chinese Academy of Sciences, Urumqi 830011, China;*
[2] *Key Laboratory of Radio Astronomy, Chinese Academy of Sciences, Nanjing, 210008, China;*
[3] *Xinjiang Key Laboratory of Radio Astrophysics, Urumqi 830011, China;*
[4] *Xinjiang Key Laboratory of Microwave Technology, Urumqi 830011, China;*
[5] *Australia Telescope National Facility, CSIRO Space and Astronomy, PO Box 76, Epping, NSW 1710, Australia;*
[6] *Department of Astronomy, Peking University, Beijing 100871, China;*
[7] *School of Astronomy and Space Science, Nanjing University, Nanjing 210093, China;*
[8] *Dr. Hans Jürgen Kärcher Engineering Office, Karben 61184, Germany;*
[9] *Northwest China Research Institute of Electronic Equipment, Xi'an 710065, China;*
[10] *OHB Digital Connect (MT Mechatronics), Mainz 55130, Germany;*
[11] *Beijing Institute of Architectural Design, Beijing 100045, China;*
[12] *Xidian University, Guangzhou Institute of Technology, Guangzhou 510555, China*

This study presents a general outline of the Qitai radio telescope (QTT) project. Qitai, the site of the telescope, is a county of Xinjiang Uygur Autonomous Region of China, located in the east Tianshan Mountains at an elevation of about 1800 m. The QTT is a fully steerable, Gregorian-type telescope with a standard parabolic main reflector of 110 m diameter. The QTT has adopted an umbrella support, homology-symmetric lightweight design. The main reflector is active so that the deformation caused by gravity can be corrected. The structural design aims to ultimately allow high-sensitivity observations from 150 MHz up to 115 GHz. To satisfy the requirements for early scientific goals, the QTT will be equipped with ultra-wideband receivers and large field-of-view multi-beam receivers. A multi-function signal-processing system based on RFSoC and GPU processor chips will be developed. These will enable the QTT to operate in pulsar, spectral line, continuum and Very Long Baseline Interferometer (VLBI) observing modes. Electromagnetic compatibility (EMC) and radio frequency interference (RFI) control techniques are adopted throughout the system design. The QTT will form a world-class observational platform for the detection of low-frequency (nanoHertz) gravitational waves through pulsar timing array (PTA) techniques, pulsar surveys, the discovery of binary black-hole systems, and exploring dark matter and the origin of life in the universe. The QTT will also play an important role in improving the Chinese and international VLBI networks, allowing high-sensitivity and high-resolution observations of the nuclei of distant galaxies and gravitational lensing systems. Deep astrometric observations will also contribute to improving the accuracy of the celestial reference frame. Potentially, the QTT will be able to support future space activities such as planetary exploration in the solar system and to contribute to the search for extraterrestrial intelligence.

*Corresponding author (email: na.wang@xao.ac.cn)

# 1 Introduction

Astronomical studies seek to reveal fundamental problems, such as the origin and evolution of the universe, the existence of dark matter and dark energy, the detection of black holes and gravitational waves, and an understanding of the nature of space-time. Research on the universe may also answer another profound question: how life came into being, and the existence of extra-terrestrial intelligence. Different frequencies of electromagnetic emission reflect the various properties of celestial objects, and multi-band astronomy provides the possibility to obtain an integrated picture of the universe. The development of radio astronomy in the 1930s opened an astonishing new window for observing the universe which complements the traditional optical band. Four major discoveries of modern astronomy: quasars, pulsars, interstellar molecules and the cosmic microwave background radiation, were all based on radio astronomy, showing its strong vitality.

It is always crucial to detect weak signals from remote objects and to study their detailed structures by enhancing the sensitivity and resolution of radio telescopes [1]. The major factor for obtaining higher sensitivity is to increase the signal collecting area. One approach is to build a larger aperture single-dish antenna. The Green Bank Telescope (GBT) [2] and the Five-hundred-meter Aperture Spherical Telescope (FAST) [3] are the most representative. Large aperture single-dish antennas are demanding in structural design, but are relatively flexible in terms of upgrading the feeds and receivers. The second approach is to build an array consisting of many small antennas. The Square Kilometer Array (SKA), for instance, is the new generation array telescope that relies on a series of new technologies in signal synthesis and processing.

To further improve the sensitivity, broad band receivers and fast data sampling are crucial. The atmospheric window for radio frequencies on Earth is 10 MHz to 1 THz (30 m to 0.3 mm)[1], with the longest observable wavelength approximately $10^5$ times the shortest ($\lambda_{max}/\lambda_{min}$), whereas in optical astronomy it is only a factor of two. Limited by the feed receiving bandwidth, radio telescopes operating in the meter-wave, centimeter-wave, millimeter-wave and even terahertz bands are developed separately. Even so, a particular telescope still needs to be equipped with receivers for multiple frequency bands. It is therefore necessary to reserve enough space for receiver installation. In the last decade, the leading technology of the ultra-wideband (UWB) receiver has significantly reduced the number of receivers compared with earlier schemes.

---

[1] The $O^2$ and $H_2O$ molecules in the atmosphere cause microwave absorption, mainly at high-frequency band, e.g., 40-70, 118 and 183 GHz, hence the small proportion of the atmospheric microwave window is not valid for ground observation.

In this study, we describe the design of the Qitai radio telescope (QTT), a 110-meter-aperture steerable radio telescope, which is proposed to be built in Qitai County of Xinjiang Uygur Autonomous Region of China. The planned observing frequencies for the QTT cover a wide range from 150 MHz to 115 GHz. To ensure high performance, especially for the higher frequency bands, the panels of the main reflector are adjustable to overcome gravity deformation. As the world's largest fully steerable radio telescope, QTT will be able to cover more than 75% of the sky down to a declination of about −40°, including a large fraction of the plane of our Milky Way [4].

The framework of this study is as follows. We briefly summarize the early science goals in sect. 2, and the site is introduced in sect. 3. The structural design, receiver mounting, data collecting and software control of QTT are described in sects. 4-7. Radio frequency interference (RFI) mitigation strategies are discussed in sect. 8. The construction plan and capabilities of QTT will be discussed in sect. 9.

# 2 Early scientific goals

The early scientific goals for the lower-frequency bands below 4 GHz are mainly aimed at pulsar studies, including using Pulsar Timing Arrays (PTAs) to detect nanoHertz gravitational waves (GW). At higher frequencies, the main science goals are star formation and evolution through detecting molecular spectral lines. In addition, the special location of the site makes the QTT unique for Very Long Baseline Interferometer (VLBI) networks.

## 2.1 Pulsars

Many, perhaps most, galaxies are believed to have a massive black hole at their center. The nanoHertz gravitational wave signal originated from the merging of two supermassive black holes in coalescing galaxies and could possibly be detected through timing observations of very stable millisecond pulsars (MSPs) forming a PTA [5]. The detection of nanoHertz GW signals is of great significance for studying the astrophysical process of merging black holes, the evolution of the early universe and the properties of GW at different frequencies [6]. The QTT equipped with a UWB receiver (0.7-4 GHz) and digital signal processor with high time and frequency resolution is a powerful tool for GW detection.

The QTT will also contribute to pulsar studies by monitoring the timing behaviors of a large sample of pulsars. Such observations provide a variety of information on the origin of timing noise, the physics of radiation, the nature of circumstellar and interstellar mediums, and the evolution of binary pulsars. Identifying and monitoring Fast Radio Bursts (FRBs) and studying their origin through polariza-tion observations is of high scientific value. Based on the pulsar population simulation adjacent to the Galaxy plane ($\leqslant \pm 5°$), Xie et al. [7] analyzed the QTT pulsar survey ability using the Phased Array Feed (PAF) receiver and predicted that about 2200 new pulsars could be discovered within $\pm 5°$ of the Galactic plane. However, this simulation includes the FAST-visible sky as well as other areas that FAST does not cover. If we only consider the non-overlap area with FAST, the predicted number of new pulsars discovered by the QTT survey would be about 680.

The pulsars that make up a PTA have rotation stability comparable to atomic frequency standards. Consequently, an important application of pulsars is to establish a new time-frequency standard, i.e., a pulsar clock, based on pulsar rotation [8]. This independent time system can be further used to detect variations in atomic time and to study the rotation of the Earth and small changes in the Solar System ephemeris. Pulsars with both radio and X-ray radiation are suitable candidates for deep space autonomous navigation. Regular timing observations by QTT will contribute to monitoring their rotation status and to updating the Pulsar Catalogue.

## 2.2 The origin of cosmic life and celestial objects

Molecular spectral lines allow us to probe the physical and chemical properties of the interstellar medium, in particular, molecular clouds and star-formation regions in the Milky Way and nearby galaxies or galaxies in the early universe [9,10]. For example, searching for known and unknown molecular line emission in star-forming regions, asymptotic giant branch (AGB) star envelopes, and planetary nebulae from the 1 cm band to the 3 mm band, is of great significance for understanding the origin of interstellar molecules and the chemical composition. The CO (1-0) line in galaxies at redshifts higher than 1.4, CS (2-1), HCN (1-0), and HCO+ (1-0) lines in galaxies at redshifts higher than 0.9 are detectable at 50 GHz (7 mm) or lower bands. Spectral lines of planetary atmospheres and comets provide crucial clues to the origin of the Solar System.

The establishment of an association between a supernova shock wave and a molecular cloud is helpful for the study of cosmic ray acceleration [11]. It is generally accepted that cosmic rays are accelerated by supernova shock waves. Whether protons, the main part of cosmic rays that produce TeV gamma-rays, are also accelerated by shock waves, the relative proportion of accelerated leptons (electrons) and hadrons (protons, etc.), and the efficiency of acceleration by supernova shock waves are still unclear. It is worth investigating the association between gamma rays and molecular clouds associated with supernova remnants, through the spectral imaging of OH masers at 1720 MHz and other spectral lines such as SiO and CS at 43.42 and 48.99 GHz.

## 2.3 Galaxies and black holes

It is generally accepted that massive black holes dwell in the center of typical galaxies. A large number of galaxies are found to be very active and are known as active galactic nuclei (AGN), manifested by strong relativistic jets and luminosity variations at radio frequencies [12,13]. The generally accepted mechanism is the accretion onto the black hole at the AGN center. The VLBI observation with sub-milliarcsecond resolution makes it possible to obtain detailed images of the relativistic jet near black holes. In particular, through the VLBI polarization measurements, we are able to study the physical mechanisms of the jet formation, acceleration and collimation processes, and reveal the density of electrons around the source and the magnetic field configuration. Higher resolution and more sensitive observations are still the key approach to revealing the interaction and co-evolution of the central black hole and its host galaxy.

In accordance with the geographical distribution, the world's major radio telescopes form the European VLBI Network (EVN), East-Asia VLBI Network (EAVN) and Chinese VLBI Network (CVN), etc. Located in the blank area of the VLBI network in the center of Eurasia, the QTT will contribute significantly to the EVN, improving the UV coverage while increasing the system sensitivity by 30% to 70% in multiple frequency bands. The QTT will also significantly improve the overall sensitivity of the CVN and EAVN. With the QTT joining the VLBI network, breakthroughs are expected in black hole and galaxy studies.

## 2.4 Dark matter

Kinematic studies of galaxies suggest the existence of dark matter halos around most galaxies. The gravitational lensing effect of the foreground dark matter halo acts on distant galaxies, providing a unique and powerful way to detect the distribution of dark matter. High-resolution and high-sensitivity VLBI observations of gravitational lensing systems can provide useful clues to the evolution of galaxies, the galaxy-dark matter relation and the distribution of dark matter in the universe [14].

The structure of HI at different redshifts and scales provides us sufficient insight into the distribution of baryonic "dark" matter, helpful for solving the problem of missing baryonic matter. An HI survey for redshift z<1 using the QTT on the galaxy cluster scale can provide intensity mapping and the two-point correlation of the detected HI halos, which will help to determine the large-scale distribution of baryonic matter.

The particle physics assumption of dark matter detection is that the dark matter may annihilate or decay, resulting in high-energy gamma rays or pairs of positive and negative cosmic ray particles. If the weakly-interacting massive particles (WIMPs), candidates for cosmic dark matter, are Ma-jorana neutralinos, their self-annihilation process occurring in the dark matter cluster will release electrons and positrons, and hence produce relativistic synchrotron radiation in the magnetic field of galaxies [15,16]. Searching for such emissions in galaxies and clusters will contribute to understanding the origin, acceleration and propagation of cosmic rays. This requires radio telescopes with very high sensitivity and wide radio frequency coverage, and the QTT may take a key role by synergizing with other telescopes in a large radio array.

## 2.5 Astrometry and reference frame

Astrometry and space positioning are based on astronomical reference frames. The radio reference frame is defined by the positions of extragalactic compact radio sources, while the optical reference frame is defined by optical stars. Optical stars that emit radio radiation are important objects to establish the connection between the radio and optical reference frames. A new version of the International Celestial Reference Frame (ICRF3) was released in 2020 [17]. It is the first celestial reference frame based on accurate position measurements from multi-frequency VLBI, including 4536 sources at S/X band (3 GHz/10 GHz), 824 sources at K band (22 GHz), and 678 sources at X/Ka band (10 GHz/33 GHz). The noise floor in individual source coordinates is at the level of 30 microarcsecond and the median uncertainty of the equatorial coordinates is improved by a factor of three compared with the previous realization, ICRF2. However, there is still room for improvement, such as increasing the number of sources at K and X/Ka bands and improving their position accuracy. This requires adding fainter reference sources with flux density down to mJy, improving their uneven distribution and selecting weaker radio calibrators at K band. In order to observe more weak sources with a high signal-to-noise ratio (S/N) and accurately determine their positions, large radio telescopes working at high radio frequencies are required.

In addition, establishing an accurate link between ICRF and Gaia Celestial Reference Frame (GCRF) is essential for fundamental astrometry in the Gaia era. VLBI astrometry of optical stars with radio emissions will significantly contribute to this. Large telescopes are much needed for accurate VLBI measurements of very weak radio stars.

## 3 The QTT site

The QTT site is located in the Eastern Tianshan Mountains, 46 km south of Qitai County, Hui Autonomous Prefecture of Changji, Xinjiang. The longitude and latitude of the site are 89°41'57" and 43°36'4" respectively, and the elevation is about 1800 m. The local terrain is a platform surrounded by mountains, with an area of 1.5 km from east to west and 2 km from south to north. The average altitude of the site is 1760 m, and the elevation of the surrounding mountains ranges from 1860 to 2250 m, giving good isolation from the external radio environment. The minimum elevation angle on the east side of the site is close to 12°, 6.5° on the south, 8° on the west and 6° on the north ridge. Preliminary geological survey and seismic safety assessment show a solid bedrock under the terrain, which ensures the requirements for the foundations of the QTT. For the antenna structural design, the seismic index is set as 9 seismic intensity without damage.

A 60-m high wind tower was built at the site, and at heights of 10, 30 and 50 m above the ground the maximum wind speed each minute is recorded. The statistical analysis of the annual data reveals that the monthly wind speed distribution of the site in autumn and winter, from September to March, is weaker than that in spring and summer from April to August [18]. Figure 1 shows that the wind speed proportion lower than 2 m/s (wind scale 2) is around 68%, lower than 4 m/s (wind scale 3) is 91%, lower than 6 m/s (wind scale 4) is 97%. By comparison, the wind speed proportion of less than 6 m/s is 56% for the GBT site in the US. For the extreme case, the proportion of wind speed greater than 17 m/s (wind scale 8) is less than 0.1% at the QTT site, and the GBT site is less than 0.5% [19]. As for the operation experience of the GBT, the QTT site satisfies the requirements ensuring the safe operation and pointing accuracy of the antenna in terms of wind load influence.

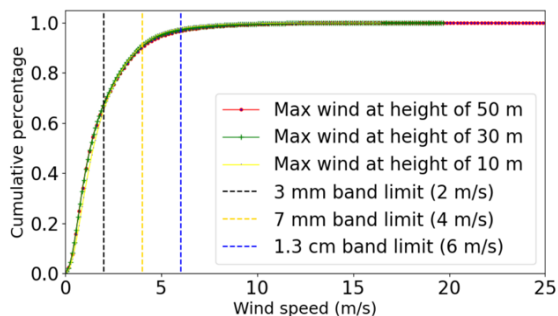

**Figure 1**  The wind speed distribution at different heights above the ground at the QTT site. The solid lines represent the maximum wind speed at heights of 10, 30 and 50 m above the ground, respectively. The dashed lines indicate the maximum wind speed limits for observations at 3 mm, 7 mm and 1.3 cm bands.

The atmosphere is not completely transparent to electromagnetic waves, especially at millimeter and sub-millimeter wavelengths. One important limiting factor for observation is the water vapor absorption caused by the atmosphere and the extra noise induced by atmospheric thermal emission [20]. Figure 2 presents the average precipitable water vapor (PWV) value of each month from 2014 to 2018 for the QTT site and the comparison with the Effelsberg and GBT sites [18]. All the data were derived from the fifth-generation reanalysis dataset of the European Centre for Medium-Range Weather Forecasting (ECMWF ERA5) products and were interpolated for the QTT site with temporal-spatial resolutions of 1 h and 0.25°×0.25°. We also installed a fully automatic CE-318 solar photometer at the site, and this is used to validate the generated PWV dataset from ECMWF ERA5. The deviation between CE-318 and ECMWF ERA5 is within 15%. Our analysis based on five-year derived data shows that the PWV content of the QTT site is 6.64 mm on average, with a mid-value of 5.11 mm. The largest value appears in July, about 14.44 mm, and the minimum is in January, about 1.95 mm. About 80% of the time the PWV level is below 10 mm. Figure 2 shows that the monthly PWV levels for the QTT site are systematically lower than those of Effelsberg and the GBT in which their annual average values are 15.98 and 16.04 mm respectively. Taking 90% as the threshold of atmosphere transmittance for 3 mm electromagnetic waves, the PWV levels are required to be lower than 5 mm in winter and 7.2 mm in summer, resulting in about 180 observable days per year at the QTT site [18]. Compared with the GBT and Effelsberg sites, the QTT site has clear advantages for high-frequency observations in millimeter band.

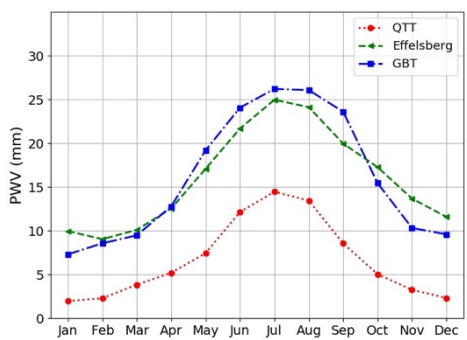

**Figure 2**  The five-year average PWVs obtained for QTT, Effelsberg and GBT sites, indicated by red, green and blue lines, respectively [18].

## 4  Antenna

### 4.1  Structure design

The QTT is a Gregorian parabolic antenna with an active main surface to correct the deformation caused by gravity. The antenna has an azimuth-elevation structure on a wheel-and-track mount. Figure 3 shows the QTT structure, including the backup structure (BUS) of the 110-m main reflector, the 12-m sub-reflector and its quadruped supporting system. The connection between the reflector back frame and antenna supporting frame is similar to the umbrella-like structure of the Effelsberg telescope in Germany [21], in which the BUS adopts a rotationally symmetric design that contains concentric radial trussed girders supported by the "umbrella" rods. The sub-reflector is connected to the truss system via the head of the quadruped supporting legs with a length of about 60 m. In order to reduce dead load and thermal deformation, the sub-reflector will be fabricated using carbon fiber-reinforced plastic.

The optical layout is a compromise between the main reflector focal ratio (0.33) and the diameter of the sub-reflector. The maximum opening angles are 74.3° and 11.0° for the feeds at primary focus and secondary focus, respectively. In order to realize multi-beam observations at both primary and secondary foci, standard paraboloids are designed for both main and sub-reflectors. And for single-beam low-frequency observations, the antenna efficiency is expected to be improved through the active shaping of the main reflector.

The receiver installation platform is one of the key issues for the QTT design. For the primary focus mode, a 20-m long prime focus platform (PFP) is integrated to carry the receivers and auxiliary equipment. There are two modes for PFP placement. In operation mode, the PFP is pulled up to the horizontal position and carries the receiver tube to the prime focus, while in stow mode, PFP is retracted to align with one of the quadrupod legs (Figure 4). For the Gregorian focus mode, a 5-m rotation turret mounted with higher frequency receivers is installed in the receiver cabin (Figure 5).

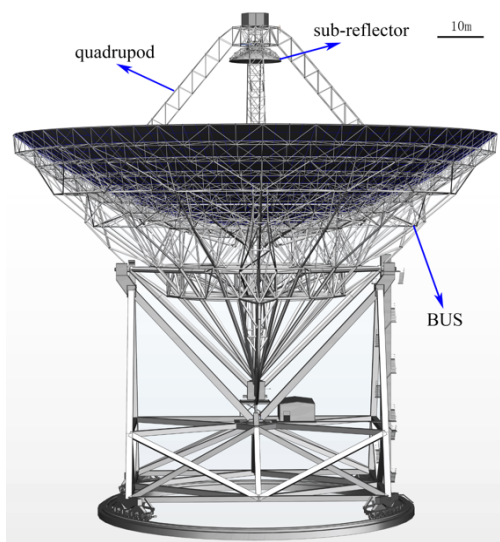

**Figure 3**  3D model of QTT antenna. The main reflector is actively adjustable through actuators.

The design for convenient and safe accessibility to the different spots of the telescope is crucial for installation and maintenance. Hence the elevator, cranes, stairs, catwalks, and platforms have been paid particular attention in the telescope structural design. Figure 6 shows parts of the access design.

The wheel-on-track system provides a stable platform for the azimuth rotation of the antenna. QTT consists of four groups of roller mechanisms, in total 32 wheels, which are installed at the bottom of the alidade frame.

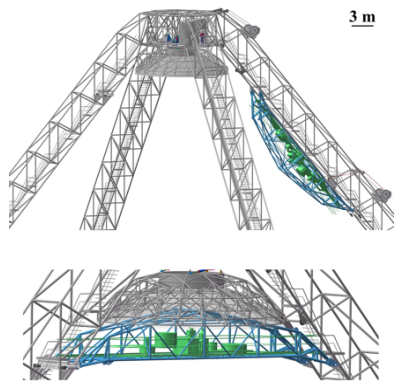

**Figure 4**  PFP in stow mode (top) and in operation mode (bottom).

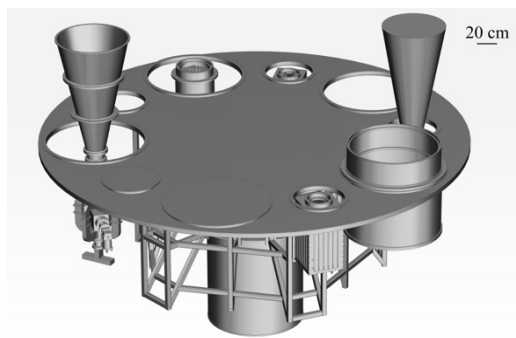

**Figure 5**  Gregorian focus feed switching mechanism, allowing for the installation of multi-band receivers.

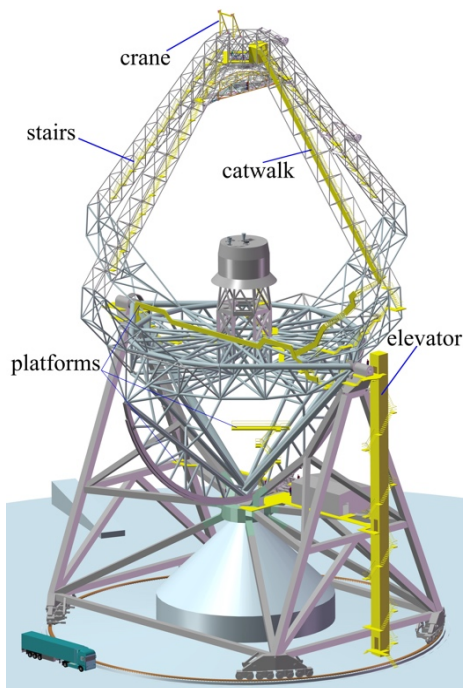

**Figure 6**  Designs for part of the access structures, indicated in yellow.

## 4.2   Deformation correction for the reflectors

The deformation of the main reflector is mainly caused by gravity, and this can be eliminated by a correction function based on the actual elevation angle [22]. Figure 7 shows the scheme of the multi-node adjustment system for the main reflector. The active main surface consists of 1920 high-precision panels, arranged on top of the upper chord of the BUS truss through the panel-adjustment mechanism known as actuators. According to a simulation analysis, the stroke range of the actuator is about 15 to 20 mm and the repeatable positioning accuracy is 15 µm. To perform 3 mm observations, the precision (rms) for the main reflector is aimed to be better than 0.2 mm under active control at the elevation angle from 80° to 10°, and for the sub-reflector, it is 0.05 mm. The panels of the main reflector are roughly in size from 4 to 6.5 m$^2$, and the accuracy of single panels should be better than 0.07 mm.

To correct the deformation caused by environmental influences such as sun exposure and steady wind load, a real-al-time photogrammetry system [23,24] consisting of a group of cameras and the targets to be imaged on the main reflector will be developed. The time scale to perform such measurement and correction is within 2 min. Because of the coefficient of thermal expansion and the heat transfer characteristics of steel, the effect of temperature unevenness on antenna structure is not in real time. According to the simulation and engineering experience, for an antenna with an aperture of 110 m, the heat transfer will take about 5 min to reach relative equilibrium, and the steady wind for the antenna is taken to be a static load. Therefore it is feasible to quickly measure and correct the deformation of the main reflector and to realize a basically closed-loop control.

Thermally-induced telescope structure deformation such as on the bearing and supporting frame can be partially corrected by temperature measurements on the telescope structure and a related "thermal model" derived from the finite element (FE) model. Inclinometers will be arranged on the alidade near the elevation bearings to recognize other unexpected distortions [25].

The position of the sub-reflector is adjustable through a heteroid Stewart-platform system. This is a modified 6-degree-of-freedom (6-DOF) parallel hexapod mechanism that provides static support to the sub-reflector as well as correcting its position and orientation through changing the length of each support arm. This allows fine adjustment to further improve the optical performance of the antenna [26] through better alignment of the main reflector and sub-reflector. The position of the sub-reflector is expected to be measured through an optical system with multiple laser beams.

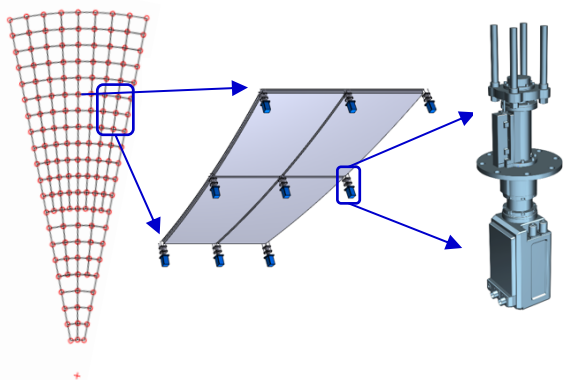

**Figure 7** The main reflector division and multi-node adjustment system. Left: actuator distribution (1/16). Middle: scheme of actuator connection with panel. Right: the actuator model.

### 4.3 Servo control and accurate pointing

The resolution of antenna is defined as proportional to $1.22\lambda/D$, where $\lambda$ is the observing wavelength, and $D$ is the aperture of the antenna. For a 110-m telescope operating at 1.3 cm, the pointing accuracy is required to be better than 5 arcsecond, and at 3 mm, ideally it should be better than 1.5 arcsecond under the condition of wind speed $\leqslant 2$ m/s and temperature drifting $\leqslant 1$ °C/h. The large aperture of QTT means it is more susceptible to environmental disturbance, such as wind, leading to more restrictive requirements on a robust servo-control system. In addition, the high inertia and low resonant frequency of the antenna increase the complexity of servo control [27,28]. To achieve the critical pointing accuracy goals, the servo control system of the antenna should have reliable repeatable dynamic performance and disturbance rejection capacity. The external disturbance and the error induced by nonlinear elements of the servo system, for example the backlash, dynamic lag and friction, need to be analyzed and simulated. Furthermore, it is necessary to make long-term synchronous tests of the servo control system, the alignment of the main and secondary reflectors, and the continual improvement of the pointing correction model. Such procedure probably will take 3 to 5 years.

The antenna servo control adopts position-loop and velocity-loop controllers based on the traditional proportional-integral-derivative (PID) algorithm. In addition, further compensation can be obtained through Disturbance Observer (DO) or Extended State Observer (ESO) to reduce the positioning error. The proportional-derivative algorithm (PD) combined with ESO is also known as Active Disturbance Rejection Control (ADRC) [29], which is probably our first approach to a better servo control. Other advanced control algorithms, e.g., $H_\infty$ [30], or Quantitative Feedback Theory (QFT) [31], are being investigated to ensure robust servo control for achieving higher precision pointing.

A typical control system consists of position and velocity loops. It is worth noting that the response of the velocity loop is faster than the position loop, and hence designing a high-performance velocity-loop controller is more effective and hence critical for improving the control performance of the servo system.

### 4.4 Wheel track and foundation

The azimuth track is connected to the foundation by anchor bolts forming a high precision and reliable platform to support the weight of the 6000-t antenna and in the meantime resist wind and earthquake loads. The wheel track of QTT will be a continuous-welded circular rail with 76 m in diameter. The base material of the track is high-strength alloy steel and the track flatness is within 0.3 mm. Due to the heavy loads and high pointing accuracy of the antenna, the uneven settlement of the track foundation must be less than 2 mm within 5 years. In order to meet this requirement, the concrete foundation consists of 59 rock-socketed piles with a diameter of 2 m and an average depth of about 23.3 m.

## 5 Receiver

### 5.1 Receiver arrangement

The receiver system selects signals of the required frequency range from the free-space electromagnetic wave, suppresses unnecessary signals or interference, and finally transmits the amplified Intermediate Frequency (IF) signal to the backend [32]. For a particular telescope, the performance of the receiver system determines the sensitivity of the system.

QTT will be equipped with broadband, ultra wide-band, multi-beam and phased array feeds low-noise cryogenic receivers from 40 cm to 3 mm bands, all linearly polarized. The current construction plan only includes receivers from 40 to 1.3 cm. The key specifications of these receivers are shown in Table 1. Among them, 40-cm, 15-cm UWB and 20-cm PAF receivers will be mounted at the primary focus and directly digitized at radio frequency (RF) without down-conversion, whereas 5, 1.3 cm broadband receivers will be mounted at the Gregorian focus and down-converted to multiple IF bands to achieve full-band acquisition.

**Table 1** Specifications of the receivers.

| Type | Band (cm) | RF Frequency (GHz) | Focus | $T_{sys}$ (K) |
|---|---|---|---|---|
| Single Beam | 40 | 0.27–1.8 | Primary | 20 |
| | 15 | 0.7–4 | Primary | 16 |
| | 5 | 4–16 | Gregory | 18 |
| | 1.3 | 16–30 | Gregory | 20 |
| PAF | 20 | 0.7–1.8 | Primary | 20 |

## 5.2 UWB Receiver

The feed is a key component for the UWB receiver. For the 15 cm UWB for instance, to achieve the bandwidth spanning 6 frequency octaves within a limited size, the feed adopts a quad-ridged horn [33]. As shown in Figure 8, the equalization characteristics of the entire working band are achieved by adding ripple grooves and medium materials to the top and middle of the horn. Probes of coaxial feedings are located at four ports at the bottom of the horn, and the dual-linear polarization is divided into three segments to output at the same time.

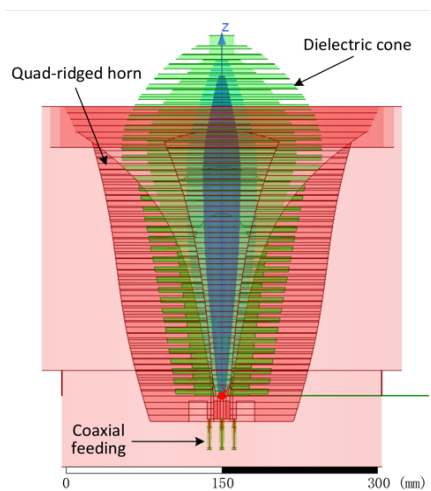

**Figure 8** Ultra-wideband quad-ridged horn feed. The part in red is the quad-ridged horn, and in green is the dielectric cone.

## 5.3 PAF

The PAF is a multi-beam receiver technology that has been widely developed in radio astronomy in recent years [34]. It uses small antennas as feeds and multiple synchronous beams are formed by electronic scanning, hence tight overlapping beams and continuous sky coverage can be achieved. Phase errors caused by feed misalignment can be compensated through appropriate phase adjustments.

The cryogenic front-end of the PAF receiver consists of 96 Vivaldi antenna units which are connected to low-noise amplifiers. The 192 signals of the dual linear polarization outputs are sampled at the receiver cabin by a signal acquisition and pre-processing unit with a 12-bit sampling rate of 2.048 giga-samples per second (GSPS). The digitized signals are then transmitted through the 100 Gigabit Ethernet (GbE) optical fiber for beam-forming and post-processing.36 beams can be formed for a maximum instantaneous bandwidth of 500 MHz selected arbitrarily from the RF passband.

## 6 Data collecting

### 6.1 The backend

To obtain high-fidelity observational signals and meet different scientific goals, a multi-function digital backend is designed with cutting-edge technologies, from which RF signals are sampled directly at the receiver frontend and processed in flexible modes. This design will effectively reduce the fluctuations of signal gain and phase caused by environmental changes during transmission. An overview of the QTT backend system is shown in Figure 9. The system is designed based on RFSoC (RF System on Chip), CPU and GPU heterogeneous architecture [35,36], including a signal acquisition and pre-processing unit, a multi-function post-processing unit, fast dump storage and a high-speed data exchange network.

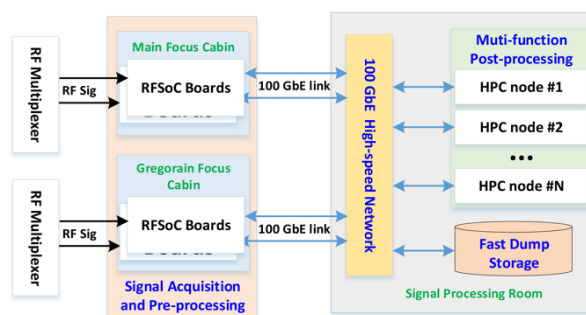

**Figure 9** A block diagram of QTT digital backend system.

The signal acquisition and pre-processing unit adopt high-performance, low-power consumption RFSoC technology with integrated high-speed ADCs. The maximum sampling rate is 4.096 GSPS with 12-bit precision. Signals from different receivers are switched through RF multiplexers. The acquisition frequency range covers the entire passband of the receivers, with a maximum simultaneous bandwidth of 14 GHz.

The multi-function post-processing unit receives the collected high-speed digital signals through a 100 GbE data exchange network, selects and loads the signal processing modes, such as pulsar mode, spectral line mode, continuum mode and baseband mode [37]. To reduce the influence of RFI, the system provides options for RFI mitigation depending on the electromagnetic environment in the observed frequency band. This unit consists of multiple high performance computer (HPC) nodes, which adjust the scale of CPU and GPU computing cores on the basis of the total signal processing bandwidth and calculation complexity, making the system highly flexible and expandable. The processed high-speed data will be buffered into the fast dump storage with a maximum rate of 4 GB/s.

### 6.2 Data archive

The QTT data management system relies on a distributed parallel file system to address the challenges of data storage and access. It will improve the aggregated bandwidth through unified service and load-balancing of multiple storage devices, and enable file system performance and capacity scalability. The key technique for the storage system is to retain data and state information, which requires the consistency of data during the data migration, automatic fault tolerance, and concurrent read/write processes.

The overall design of the QTT data management system is shown in Figure 10. The design consists of three tiers. Tier 0 is responsible for online archiving from the observational data acquisition system. Data will be pre-processed and written into a temporary storage server. After the validity check, the data are distributed to the permanent onsite archive. Tier 1 manages the remote backup of original data and the metadata extraction, this is synchronized with the onsite archiving through a dedicated network. The metadata extracted from the raw data will be imported into the corresponding database for data release. Tier 2 handles the publication and query facilities of the raw data. It will be built according to the latest International Virtual Observatory Alliance (IVOA) standards. Users will be allowed to access the online data retrieval platform through web browsers, standard virtual observatory (VO) tools, various scripts, etc.

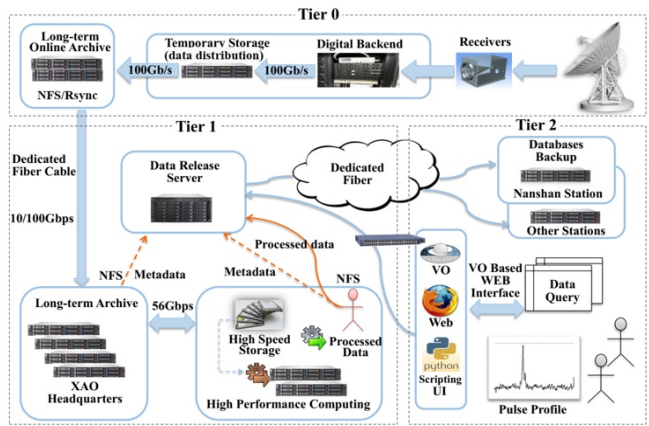

**Figure 10** Overall design of the QTT data management.

### 7 Software control center

With the development of software, sensor and network technology, the astronomical observation and control system has developed from traditional automatic control to flexible remote control [38]. The software control center of QTT is designed to adopt a modular distributed microservice framework [39]. Each module provides an independent service interface through Remote Procedure Calls (RPC) and message queues in microservice mode. Microservice and modular design can reduce the coupling between modules, improve the flexibility and scalability of the system, and better meet the needs of future equipment and scientific upgrading. The architecture of the QTT software control center is shown in Figure 11. The user interface will be developed through web technology. Users can observe, control or view the telescope status through an arbitrary web browser remotely.

The main functions of the software control center include antenna driving, feed switching, receiver and backend configuration and system monitoring. The main reflector is actively adjusted through actuators according to the feedback of the photogrammetric system. The environmental conditions are collected via meteorological and electromagnetic monitoring systems which provide criteria for safe operation and dynamic scheduling. Scientific observation modes mainly include pulsar, spectral line, continuum and VLBI. The system will rapidly measure the telescope efficiency and gain of each band and conduct the telescope pointing correction. The system monitoring module mainly provides updated information for vulnerable hardware or fatigue-prone devices, to prompt early warning and reduce the impact of device failures and potential safety hazards.

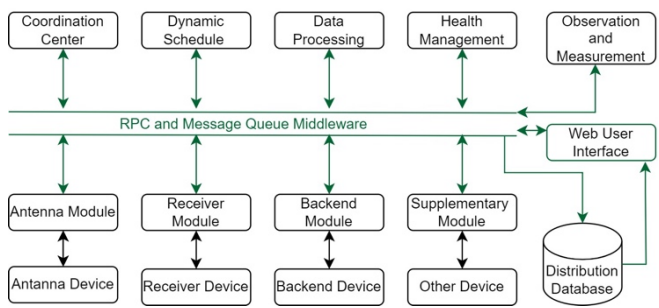

**Figure 11** Architecture of QTT software control center.

### 8 RFI mitigation

The QTT site is in a basin surrounded by mountains, and the terrain provides natural shielding which blocks the external radio interference to a great extent. The QTT site has been specially selected to be as free of external terrestrially generated RFI as possible over a large radio frequency range. The site will be maintained as such and will be established as a radio quiet zone (RQZ) in the future. The RQZ is an area with a radius of 30 km, which consists of the core zone, restricted zone and coordination zone from inside to outside with reducing control levels. The use of radio transmitting equipment and electronic products or facilities that generate electromagnetic radiation in RQZ shall be subject to the RQZ protection measures.

RFI is usually caused by mobile communications, radio and digital television, satellite communications, and radio navigation systems, and mainly appears in the low-frequency band, i.e., below 3 GHz. Large telescopes are very sensitive to the signals radiated by various electronic components. The radio telescope itself is an integration of a large number of electronic components, and their radiation coupled with outside RFI makes the electromagnetic environment of the telescope very complicated. Hence the electromagnetic compatibility of the telescope has an important impact on the performance of the system. In order to limit the influence of RFI to a minimum, electromagnetic compatibility (EMC) design according to the ITU-R RA.769 recommendation[2] is formulated by setting thresholds of allowable radiation levels for equipment located on QTT sites [40].

Multi-layer shielding is a further approach to suppress RFI generated by various electronics at the site. In particular, most of the electronics will be sheltered in a high-performance shielding chamber in the basement under the telescope. Otherwise, the electronic components distributed in the laboratory are arranged in shielded cabinets or boxes. The buildings at the site are shielded to reduce the RFI generated by lamps, switches, etc. Earlier experiments show the shielding efficiency of this scheme is about 30 dB at 0.1-1 GHz [41].

In addition, a dedicated reference antenna will be installed for monitoring and further mitigating the RFI during the data sampling process in future observations. The digital backend system also provides the possibility of adaptive RFI filtering.

## 9 Discussion and conclusions

The construction scheme of the QTT is closely coordinated with the early science goals. The operation includes pulsar, spectral line, continuum and VLBI observational modes. The QTT will be equipped with UWB receivers and a PAF. A heterogeneous system based on an RFSoC, CPU and GPU cluster realizes the ultra-wideband RF signal sampling and data processing.

The overall design of the antenna adopts a lightweight and shape-preserving structure. An active main reflector and adjustable secondary reflector further ensure the efficiency of system. The main structure of QTT will take about five years to build, after which low-frequency receivers will be installed, and the project will enter the system testing and commissioning phase. After continuously improving the correction model for the main reflector surface, sub-reflector positioning and pointing accuracy, higher frequency observations at 7 and 3 mm will be started.

Currently, there are only two 100-meter fully steerable radio telescopes operating in the world, the Effelsberg telescope [21] and GBT [2]. As introduced in sect. 4, the QTT adopts symmetric structure via umbrella-like supporting frame. The advantage of such design is that the weight of the telescope can be significantly reduced, while the disadvantage is the sub-reflector inevitably blocking the main reflector, resulting in a reduction in telescope efficiency. The GBT gave priority to avoiding aperture blockage, hence an asymmetric structure was designed in which the sub-reflector is off-axis arranged relative to the paraboloid main reflector. The trade-off of such a design is the increasing complexity of the structure of the supporting truss and frame, and the weight of the telescope is hard to control. Other than that, QTT and GBT can both operate at frequency bands up to 3 mm, and the high precision of the active main reflector is realized through actuators.

At 3 mm wavelength the ALMA [42] is the most powerful interferometer with an unprecedented resolution of 0.04 arcsecond at the longest baseline, and roughly equivalent sensitivity to a hundred-meter single dish. The larger beam of the QTT will enable us to conduct extensive and fast pilot surveys with limited resolution, which can guide follow-up high-resolution observations with ALMA. Furthermore, while the interferometer is good at zooming into the compact emission regions, it also loses information about the large-scale diffuse emission. The QTT will be able to fill in this gap. Combining observations with the interferometer with those of the QTT will allow a more complete determination of the appearance of the sky from small to large scales.

The entire radio frequency band covers from meter to submillimeter wavelengths. Operating at 70 MHz-3 GHz, FAST has extremely high sensitivity by forming a 300-meter aperture instantaneous paraboloid in the observation direction by active control [3]. The Commensal Radio Astronomy FAST Survey (CRAFTS) [43] and the Galactic Plane Pulsar Snapshot (GPPS) [44] survey have now found more than 600 pulsars, and its powerful detection capabilities could lead to the discovery of rare objects such as black holes and more unknown phenomena in the future [45]. The follow-up observation and verification of the new objects needs a joint effort with QTT, the largest steerable telescope with tracking ability [46]. As the QTT has much wider frequency coverage, coordinated observations with FAST can be made to investigate the frequency-related physics properties.

Real-time processing and storage of massive data quantities could be challenging in the future. Taking pulsar observations as an example, in order to correct for interstellar medium propagation effects, pulsar observation data require extremely high time and frequency resolution and sampling intervals of tens of microseconds. Some special high-precision baseband observation systems require Nyquist rate sampling in the full band of several GHz, which requires extremely high rates of online data processing.

*This work was supported by the National Key Research and Development Program of China (Grant Nos. 2021YFC2203501, 2021YFC2203502, 2021YFC2203503, and 2021YFC2203600), the National Natural Science Foundation of China (Grant Nos. 12173077, 11873082, 11803080, and 12003062), the Scientific Instrument Developing Project of the Chinese Academy of Sciences (Grant No. PTYQ2022YZZD01), the Operation, Maintenance and Upgrading Fund for Astronomical Telescopes and Facility Instruments, budgeted from the Ministry of Finance of China and Administrated by the Chinese Academy of Sciences, and the Chinese Academy of Sciences "Light of West China" Program (Grant No. 2021-XBQNXZ-030).*